\pdfoutput=1
\documentclass{JINST}
\usepackage{textcomp} 
\DeclareGraphicsExtensions{.pdf,.png,.jpg} 

\title{Development and validation of a 64 channel front end ASIC for 3D directional detection for MIMAC}

\author{ J.P.~Richer$^a$, O. Bourrion$^a$\thanks{Corresponding author.}, 
G.~Bosson$^a$, O.~Guillaudin$^a$, F.~Mayet$^a$, D.~Santos$^a$.\\
\llap{$^a$}Laboratoire de Physique Subatomique et de Cosmologie,\\ 
Universit\'e Joseph Fourier Grenoble 1,\\
  CNRS/IN2P3, Institut Polytechnique de Grenoble,\\
  53, rue des Martyrs, Grenoble, France\\
  E-mail: \email{olivier.bourrion@lpsc.in2p3.fr}}

\abstract{A front end ASIC has been designed to equip the \textmu TPC prototype developed for the MIMAC project, which requires 3D reconstruction of  low energy particle tracks in order to perform directional detection of galactic Dark Matter.
Each ASIC is able to monitor 64 strips of pixels and provides the ``Time Over Threshold'' information for each of those. These 64 digital informations, sampled at a rate of 50\,MHz, can be transferred at 400\,MHz by eight LVDS serial links. Eight ASIC were validated on a $2 \times 256$ strips of pixels prototype.}


\begin{document}

\section{Introduction}
Directional detection of dark matter is known to be a promising search strategy of galactic Dark Matter \cite{Spergel,Ahlen}). Recent studies have shown that, within the framework of dedicated statistical data analysis, a low exposure directional detector could lead either to a high significance discovery of galactic Dark Matter \cite{billard.2010a,billard.2011} or to a conclusive exclusion \cite{billard.2010b}.
A gaseous micro-TPC matrix, filled with either $\rm ^3He$, $\rm CF_4$ or $\rm C_4H_{10}$ has been developed within the MIcro TPC MAtrix of Chambers (MIMAC) project \cite{MIMAC}.
To demonstrate the relevance of the concept, a specific front-end ASIC and a dedicated acquisition electronic 
were developed in order to equip a prototype detector featuring an anode of $\rm 10.85 \times 10.85\ cm^2$ where 
$2 \times 256$ strips are monitored. 
The ASIC is designed to allow an auto-triggered acquisition electronic which uses embedded processing to reduce data transfer to its useful part only, i.e. decoded coordinates of hit tracks and corresponding energy measurements.

\section{MIMAC detector readout principle}
\label{detPrinciple}
\begin{figure}[ht]
\begin{center}
\includegraphics[width=0.6\textwidth]{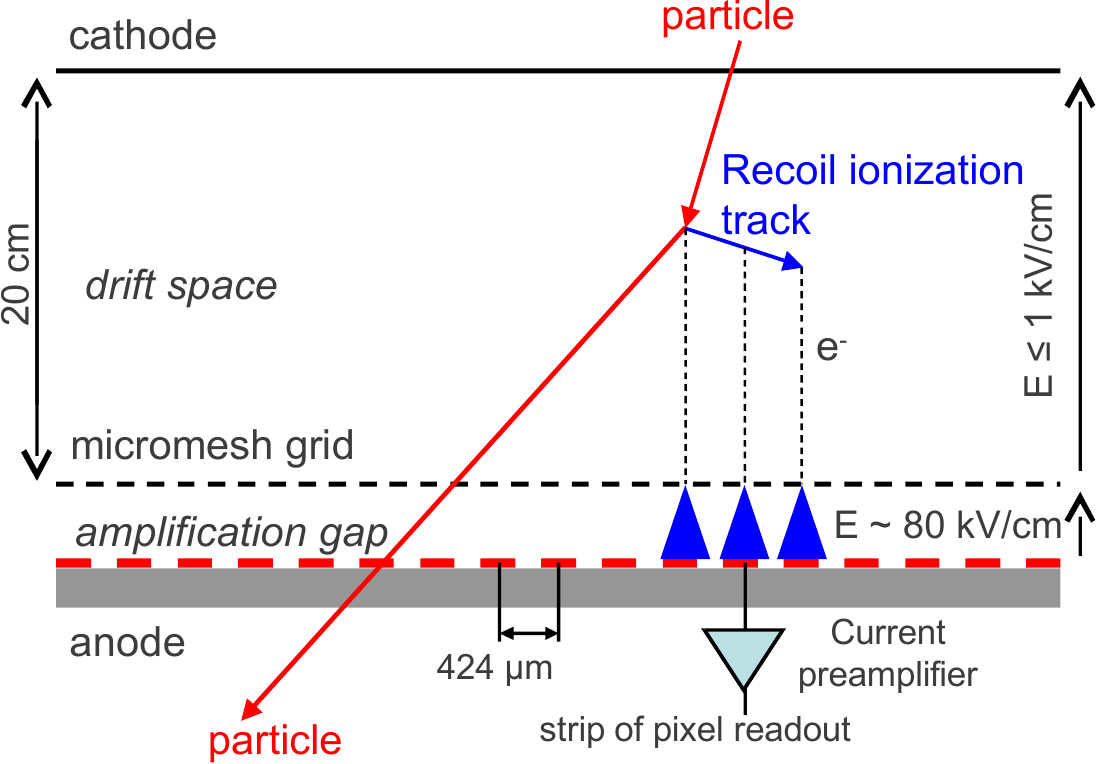}
\caption{Schematic of the MIMAC micro-TPC using a micromegas composed of a pixelized anode featuring two orthogonal series of 256 strips of pixels  and a micromesh grid defining the delimitation between the amplification (grid to anode) and the drift space (cathode to grid).}
\label{microTPC}
\end{center}
\end{figure}

\begin{figure}[ht]
\begin{center}
\includegraphics[width=0.8\textwidth]{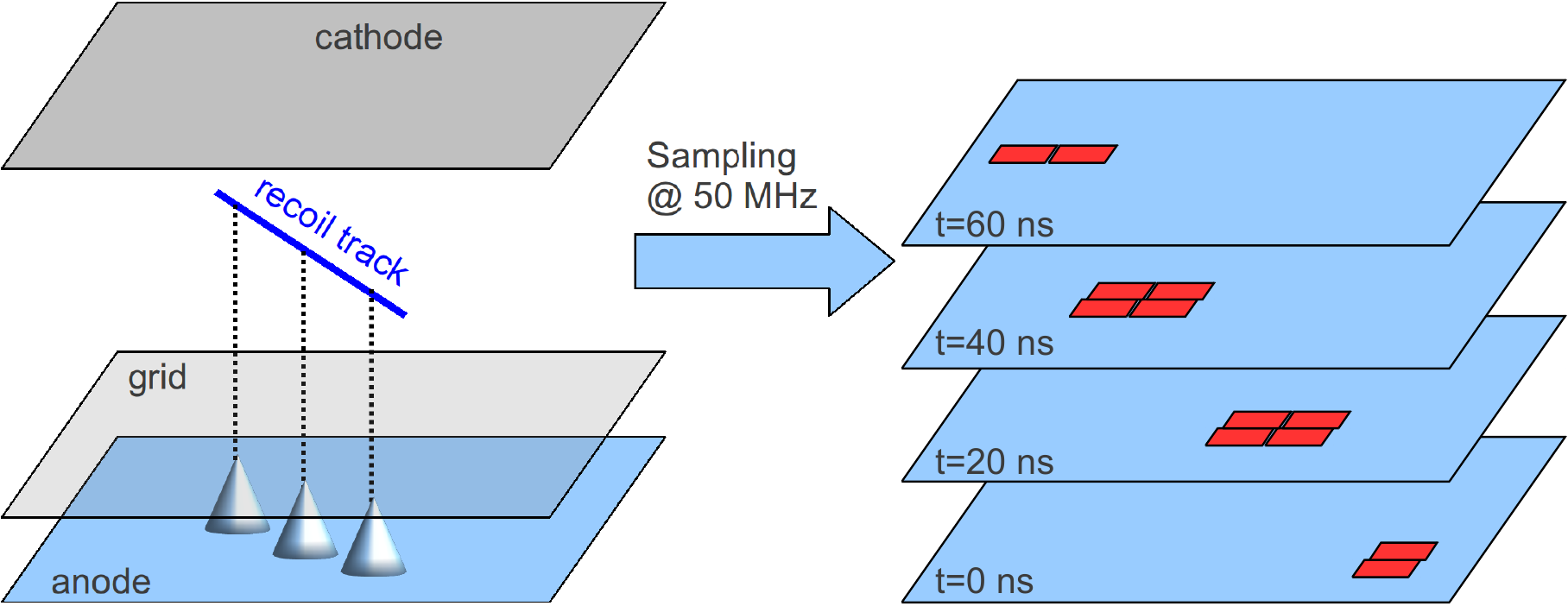}
\caption{The coordinates in the anode plane (X, Y) are reconstructed by collecting primary electrons produced in the drift region. Knowing the electron drift velocity, the third dimension (Z) is obtained by sampling  the anode signal every 20\,ns. Due to the multiplexed readout of the anode, each time slice picture is rectangular.}
\label{ThirdDim}
\end{center}
\end{figure}

As shown in figure~\ref{microTPC}, the MIMAC prototype \textmu TPC is composed of a pixelized anode featuring two orthogonal series of 256 strips of pixels (X and Y) \cite{Iguaz} and a micromesh grid defining the delimitation between the amplification (grid to anode) and the drift space (cathode to grid).
Each strip of pixels is monitored by a current preamplifier and the fired pixel coordinate is obtained by using the coincidence between the X and Y strips (the pixel pitch is 424\,\textmu m). A coincidence is defined as having at least one strip of pixels fired in each direction (X, Y) at the same sampling time.  The ionization energy of the recoil particle is obtained by instrumenting the micromesh grid with a Charge Sensitive Preamplifier (CSP).
As illustrated in figure~\ref{ThirdDim}, the coordinates in the anode plane (X, Y) are reconstructed by collecting primary electrons produced in the drift region. Knowing the electron drift velocity, the third dimension (Z) is obtained by sampling  the anode signal every 20\,ns. Note, that due to the multiplexed readout of the anode, each time slice picture is rectangular.

\section{Front end ASIC}
\label{ASIC.sec}
\subsection{Requirements}
In the early stage of the project it was decided to design an ASIC in order to be able to fulfill the final objective, which is to equip about 2500 chambers of 1024 strips of pixels (512+512). This minimizes space requirement and power demand while allowing cost reduction on a large scale. After going through a first prototype phase, 16 channels ASICs equipping a $2 \times 96$ strips of pixels chamber \cite{Richer}, a 64 channel version was designed. This was determined to be a good balance between integration scale on one side and complexity, fabrication yield and available packages on the other side. 
To be able to recover the third coordinate (Z) of the track, a fast switching current comparator having a threshold as low as 200\,nA must be designed in order to have a precise time over threshold measurement of each current preamplifier output. This requirement is driven by the worst case where the recoil energy is as low as 500\,eV in a chamber having its gain  limited to 3000 and where the diffusion is maximized (i.e interaction at the farthest of the anode) and the recoil track is parallel to the anode (the charge deposit is distributed along different strips). Another strong system requirement is to minimize the board level interconnection to allow an easy integration with a readout system.

\subsection{Design overview}
\begin{figure}[ht]
 \begin{center}
 \includegraphics[width=0.75\textwidth]{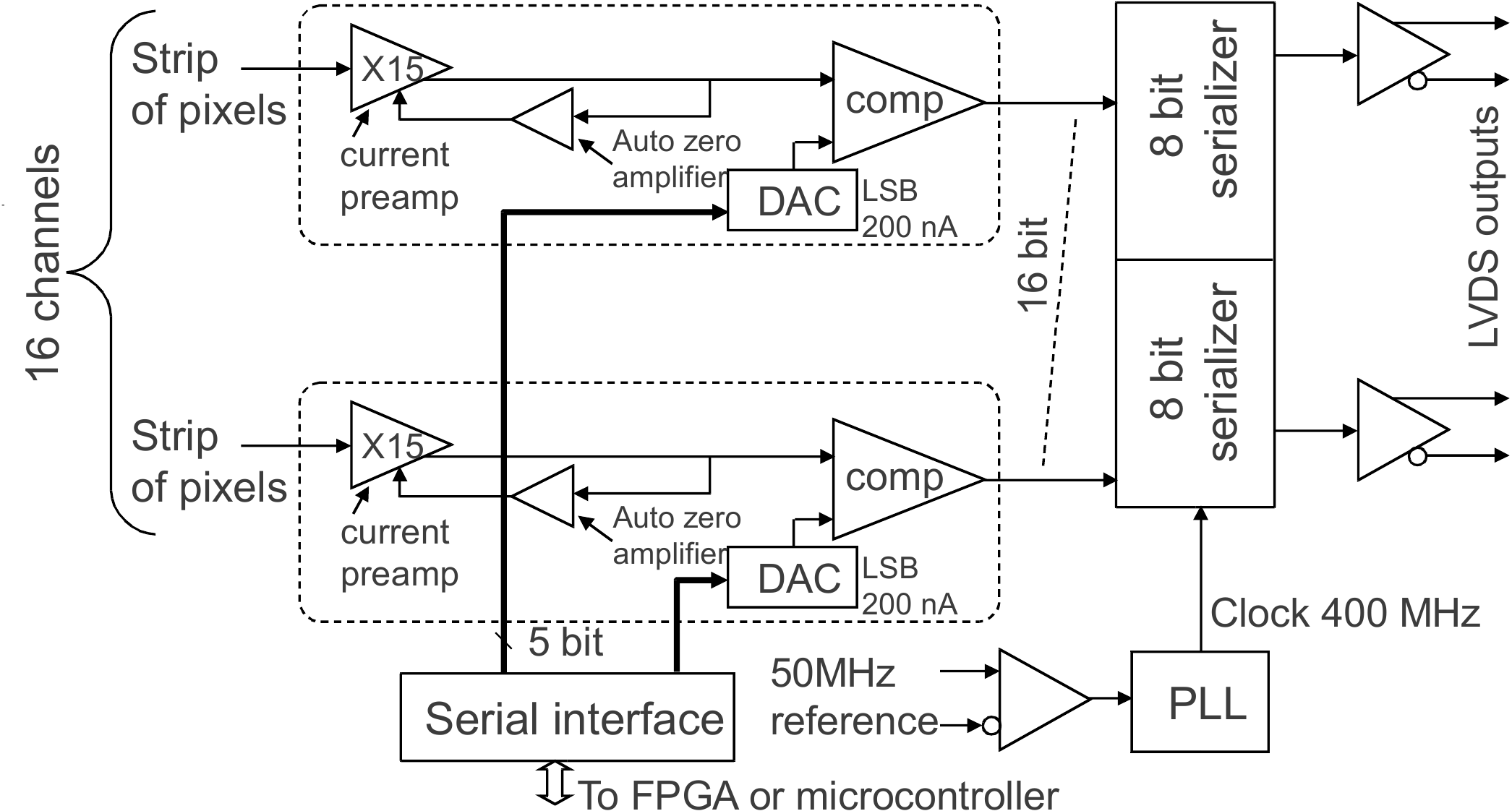}
\caption{Front end ASIC block diagram. The 64 channels are decomposed in four groups of 16 channels. Each channel is composed of a current preamplifier having a gain of 15, a fast comparator and a 5 bit DAC for setting the threshold. A serial link is used to configure the DAC. The comparator outputs are transferred serially.}
\label{BlockASIC}
\end{center}
\end{figure}

Figure~\ref{BlockASIC} shows the block diagram of the front-end ASIC; it is composed of four groups of 16 channels. Each channel is composed of a current preamplifier having a gain of 15, a fast comparator and a 5 bit DAC for setting the threshold. 
A trade-off was made between the DAC resolution and the achievable preamplifier offset. 
The DAC dynamic range, and thus its design complexity, can be greatly reduced by using an autozero preamplifier. This kind of amplifier measures periodically (every second) its offset during a few dozen \textmu s and determines the compensation to apply to reduce the residual output offset. 
The 5 bit DAC is designed with a 200\,nA LSB (input equivalent: 13.3\,nA).

The comparator outputs are sampled at a 50\,MHz rate and serialized at 400\,MHz, thereby reducing the interconnection by a factor of eight and diminishing the power consumption.  The serial outputs rely on the Low Voltage Differential Signaling (LVDS) standard to lower the electronic noise. It should be noted that using the same reference clock allows synchronous sampling between ASICs.
A slow serial link is implemented to allow ASIC configuration. With this link, it is possible to individually adjust the thresholds (DAC settings) and to enable/disable each channel (kill eventual dead channels, ...). Finally, the fixed training pattern is also provided via this link. It is used for the synchronization of the digital readout electronics with the high speed serial links.

\subsection{Input stage}
\begin{figure}[ht]
 \begin{center}
 \includegraphics[angle=-90,width=0.65\textwidth]{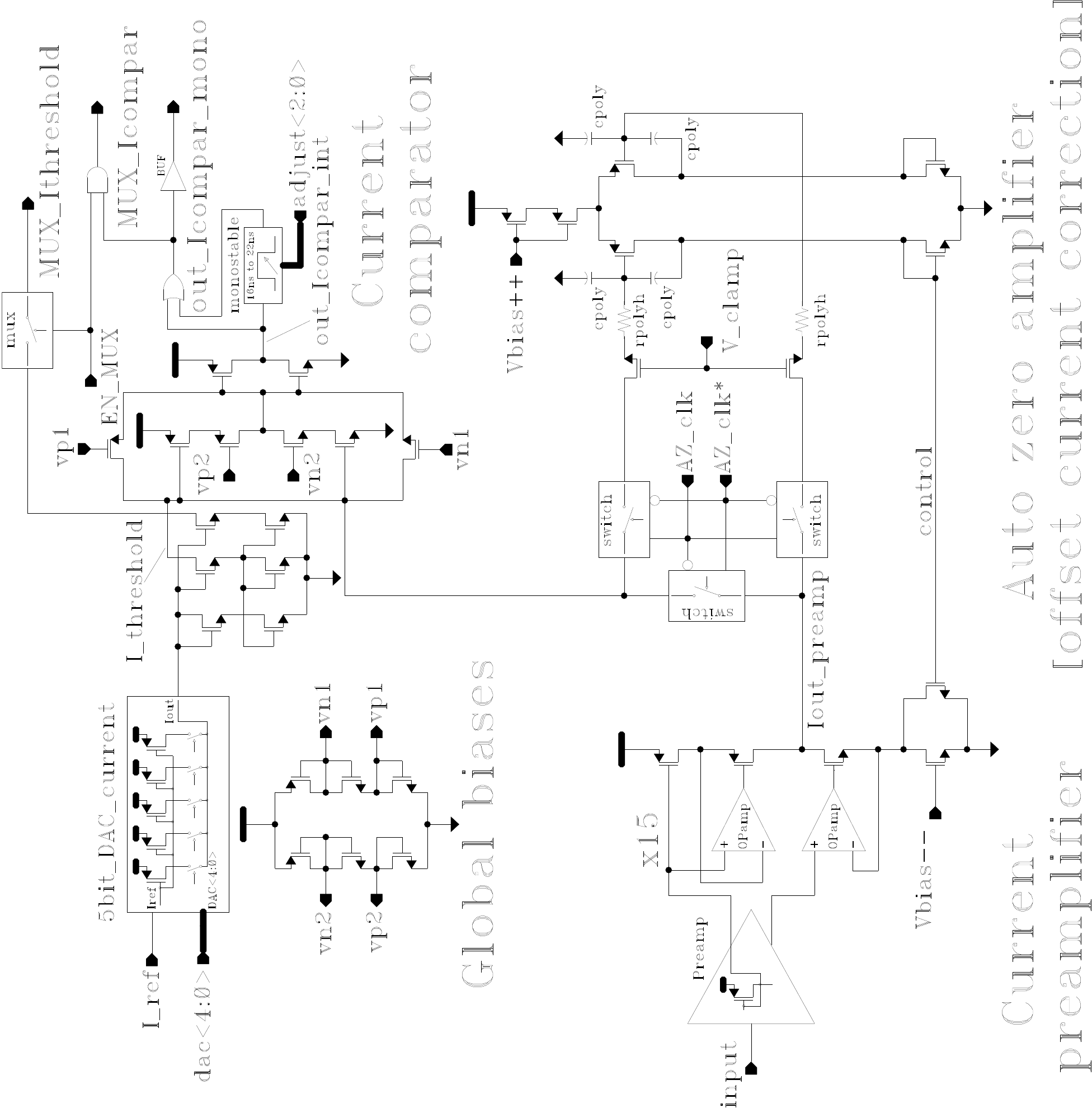}
\caption{Input stage schematic}
\label{InputStageFig}
\end{center}
\end{figure}

Originally an energy measurement per group of sixteen channels (summing and shaping) has been implemented in the first ASIC version \cite{Richer}. Experimental results with the MIMAC detector showed that this strategy is not efficient for low energy tracks, thus an instrumentation of the micro-mesh grid with a Charge Sensitive Preamplifier is used to obtain the total energy signal. Due to this initial choice, the input stage (figure~\ref{InputStageFig}) consists of a low noise charge preamplifier based on a folded cascode structure. The integration current, flowing through the capacitor, is copied and amplified with a gain of 15 to generate the input signal of the current comparator.

This current comparator \cite{linan}, used here as an amplifier, consists of a CMOS inverter equipped with two cascode transistors used to decrease the influence of the input to output parasitic capacitor.
Also, for improving the comparator commutation speed, the CMOS inverter is kept in linear mode by the use of the two feedback transistors barely maintained ``OFF'' in the quiescent state (no input current, no threshold applied). A common block of voltage dividers  per group of 16 channels (figure~\ref{InputStageFig}), generates the four biasing voltages ``vnx'' and ``vpx''. In this common block, the transistors geometries ratios were determined by simulation to obtain the optimal biasing values. 

\begin{figure}[ht]
 \begin{center}
 \includegraphics[width=\textwidth]{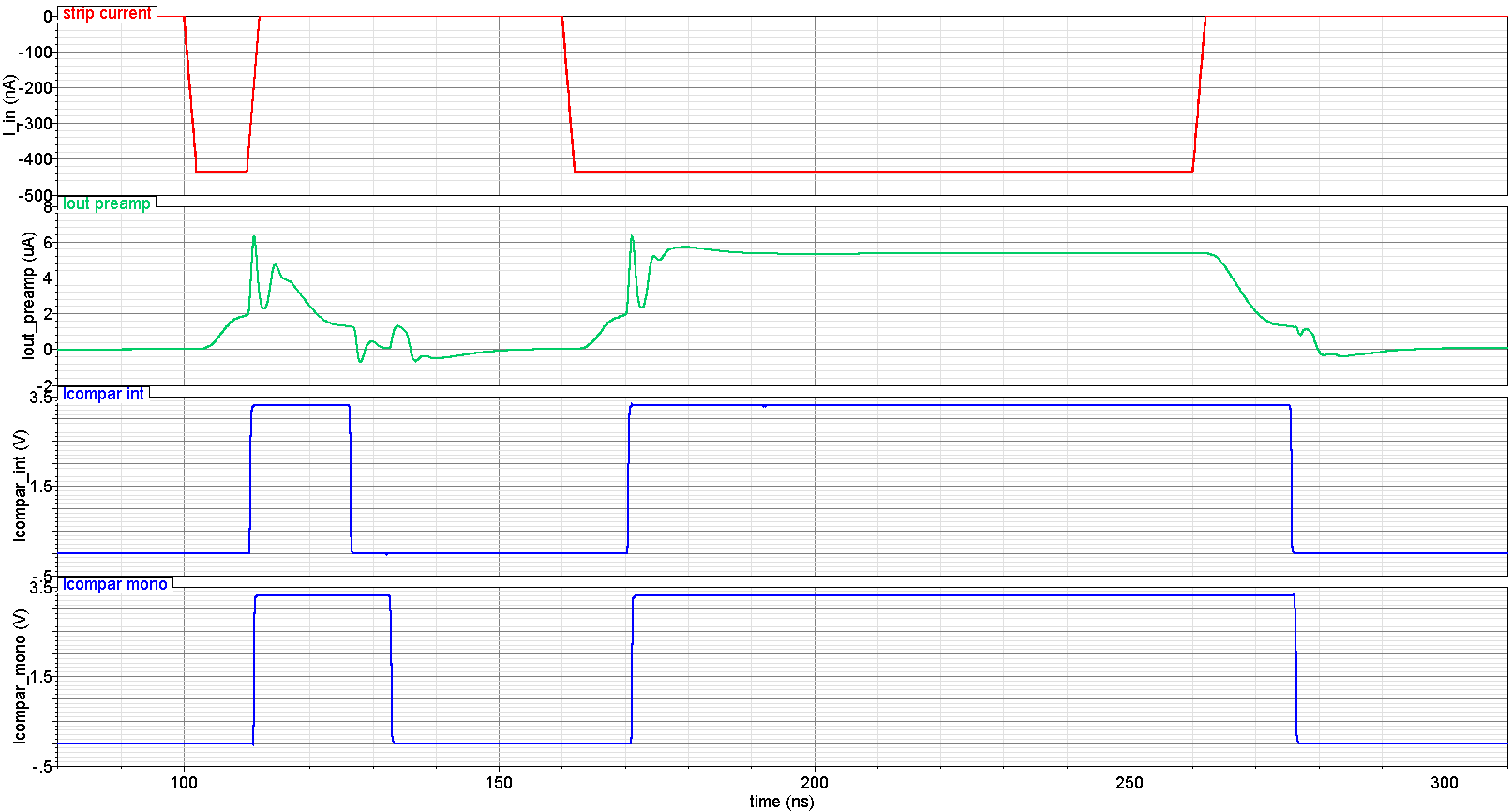}
\caption{Simulation result showing the fast comparator commutation and the pulse lengthening for input current lasting less than 20\,ns. First curve shows the preamplifier input current, second curve shows the preamplifier output current. The two bottom curves show respectively the current comparator and OR gate outputs.}
\label{MonostableFig}
\end{center}
\end{figure}

When a threshold current (provided by the 5 bit DAC) is applied (see  node ``I\_threshold'' in figure~\ref{InputStageFig}), one of the two feedback transistors starts conducting and therefore allows the current flow. Then, when an input current higher than the current threshold is applied, the previously conducting transistor is switched ``OFF'', the CMOS inverter toggles and the other feedback transistor starts conducting. The comparator therefore delivers a digital signal whose duration is equal to the current duration.
In order to cope with the short duration signals ($<$20\,ns), the channel output signal is the output of an OR gate which is fed by the comparator output itself and by a programmable monostable triggered by the same comparator output signal (figure~\ref{MonostableFig}).

\begin{figure}[ht]
 \begin{center}
  \includegraphics[width=0.48\textwidth]{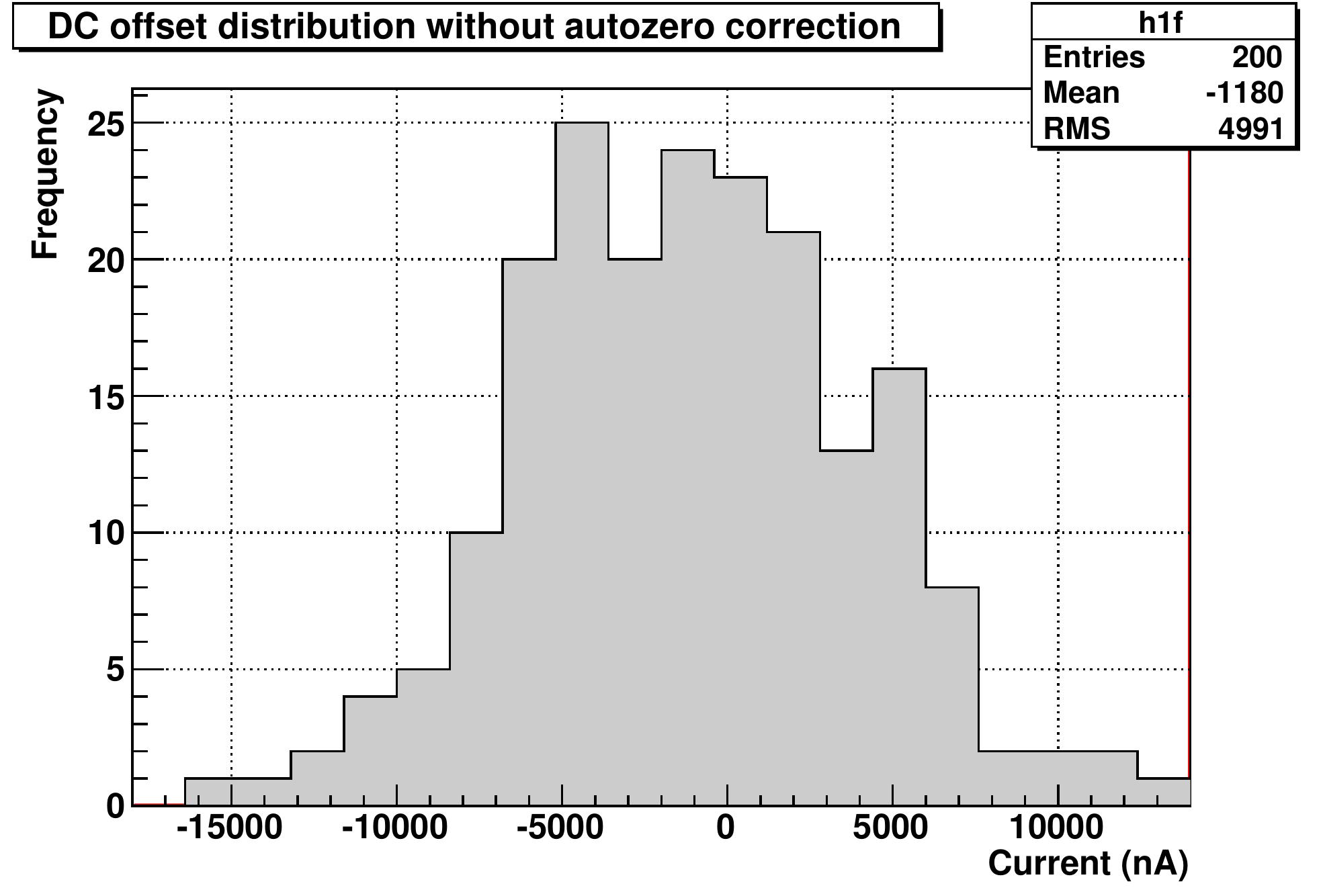}
  \includegraphics[width=0.48\textwidth]{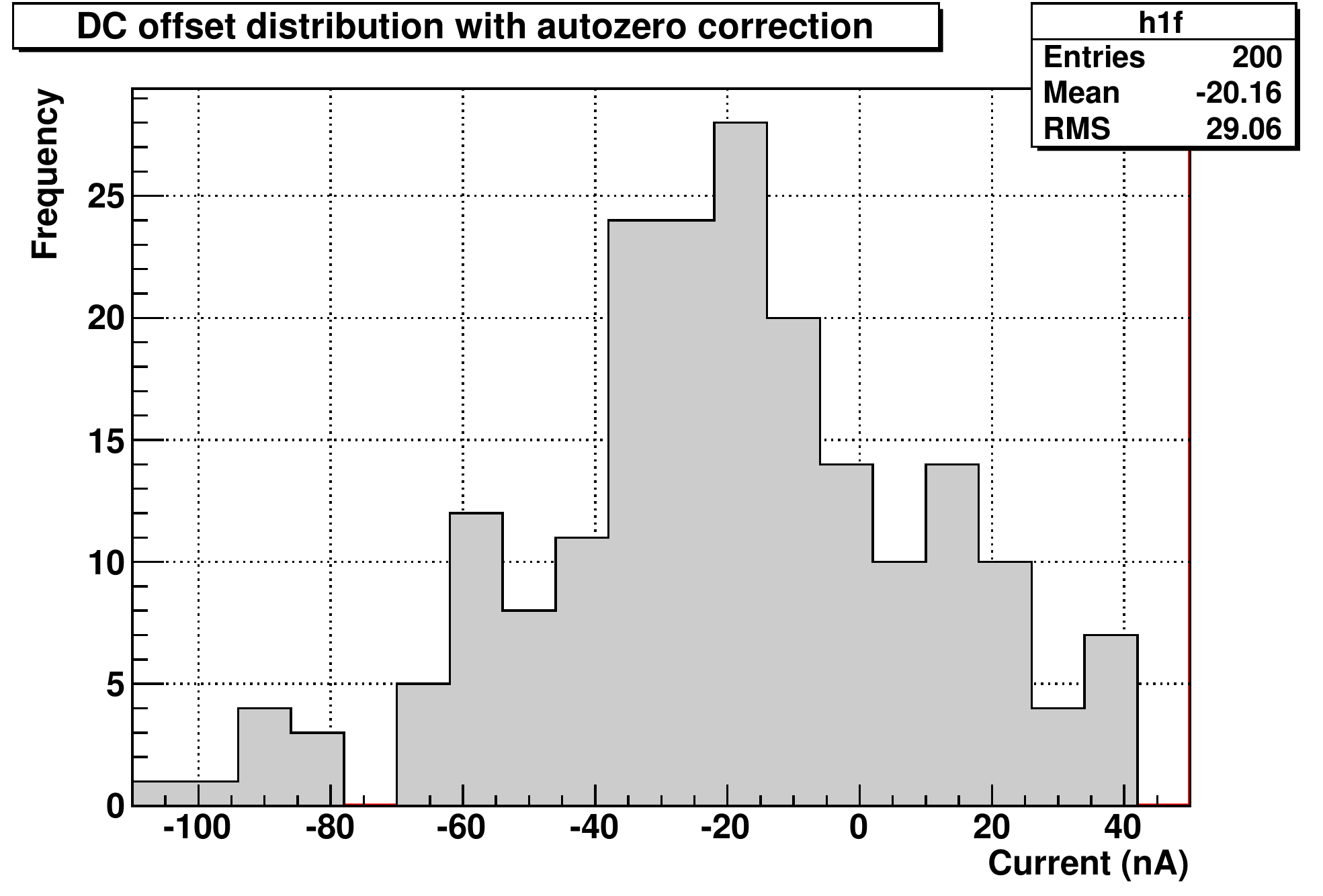}
\caption{Offset current distributions without and with auto-zero correction.}
\label{AZResults}
\end{center}
\end{figure}

In order to be able to use this reduced resolution DAC, an auto-zero amplifier has been added for minimizing the DC offset current in the output branch of the preamplifier, see results in figure~\ref{AZResults}. This offset is essentially due to transistor mismatches. At a slow rate (1\,Hz) the current output of the preamplifier is disconnected from the discriminator and connected to one of the differential amplifier inputs. The offset current charges or discharges a capacitor and the amplifier compares the voltage capacitor to the DC input voltage of the current comparator. The offset correction is applied to the output branch via a current mirror and the ``control'' line. During the auto-zero DC correction, a certain isolation to the input signal has to be implemented in order to provide adequate offset cancellation. For this, series resistors and long channel transistors have been added. The gate voltage ``V\_ clamp'', which is accessible outside of the ASIC, provides an adjustment of the filtering time constant. 

\subsection{Serializer}
\begin{figure}[th!]
\begin{center}
\includegraphics[width=0.8\textwidth]{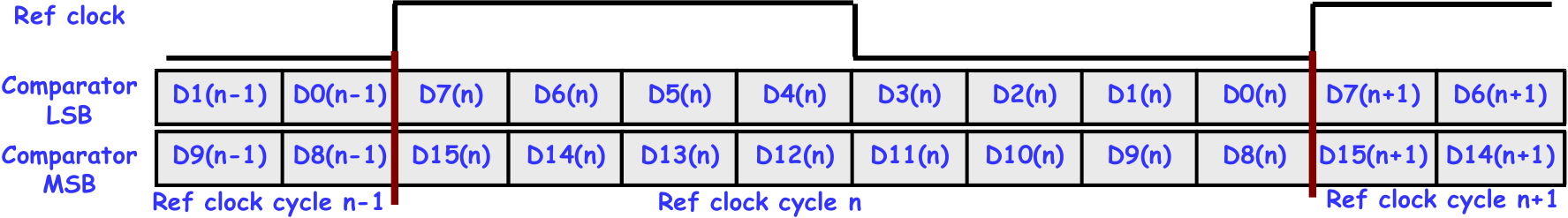}
\caption{Content of the two serial frames related to a group of sixteen channels.}
\label{frame}
\end{center}
\end{figure}

One major difficulty of this front end ASIC is to get all the comparator outputs out of the chip. This is a concern for several reasons. First, the interconnections have to be kept to a minimum in order to equip large detectors and second, the digital noise level induced by these outputs has to be kept low. This last reason leads to the choice of LVDS differential outputs. 
The problem is that the LVDS signaling requires two wires per connection and a relatively high power. The remedy to this is to implement a serializer that will work eight times faster and thus time multiplex data into a single differential pair. Instead of providing the eight differential data at a rate of 50\,MHz, the data can be transferred on a single pair at a rate of 400\,MHz, as shown in figure~\ref{frame}. 
The 400\,MHz clock is generated by a Phase Locked Loop (PLL) circuit synchronized with the 50\,MHz reference clock. Using the same reference clock allows synchronous sampling between ASICs. 
This ``classical'' PLL structure  \cite {dzahini} is composed of four blocks (VCO, charge pump, divider and phase detector). The Voltage Control Oscillator (VCO) consists of a ring oscillator with seven ``starved'' inverters whose delays are controlled by the charge pump delivered voltage. This charge pump is driven by a phase detector used to compare the phase of the VCO output divided by eight with the reference clock. Due to the large capacitors required, the charge pump filter is implemented externally.
The LVDS transmitter and receiver are based on structures described in \cite {boni}. For testing and synchronization purposes, a multiplexer is implemented just before the serializer. This enables the user to select either the sending of a known fixed pattern or of the comparator outputs. 

\section{ASIC Performances}
\begin{figure}[th!]
\begin{center}
 \includegraphics[width=0.48\textwidth]{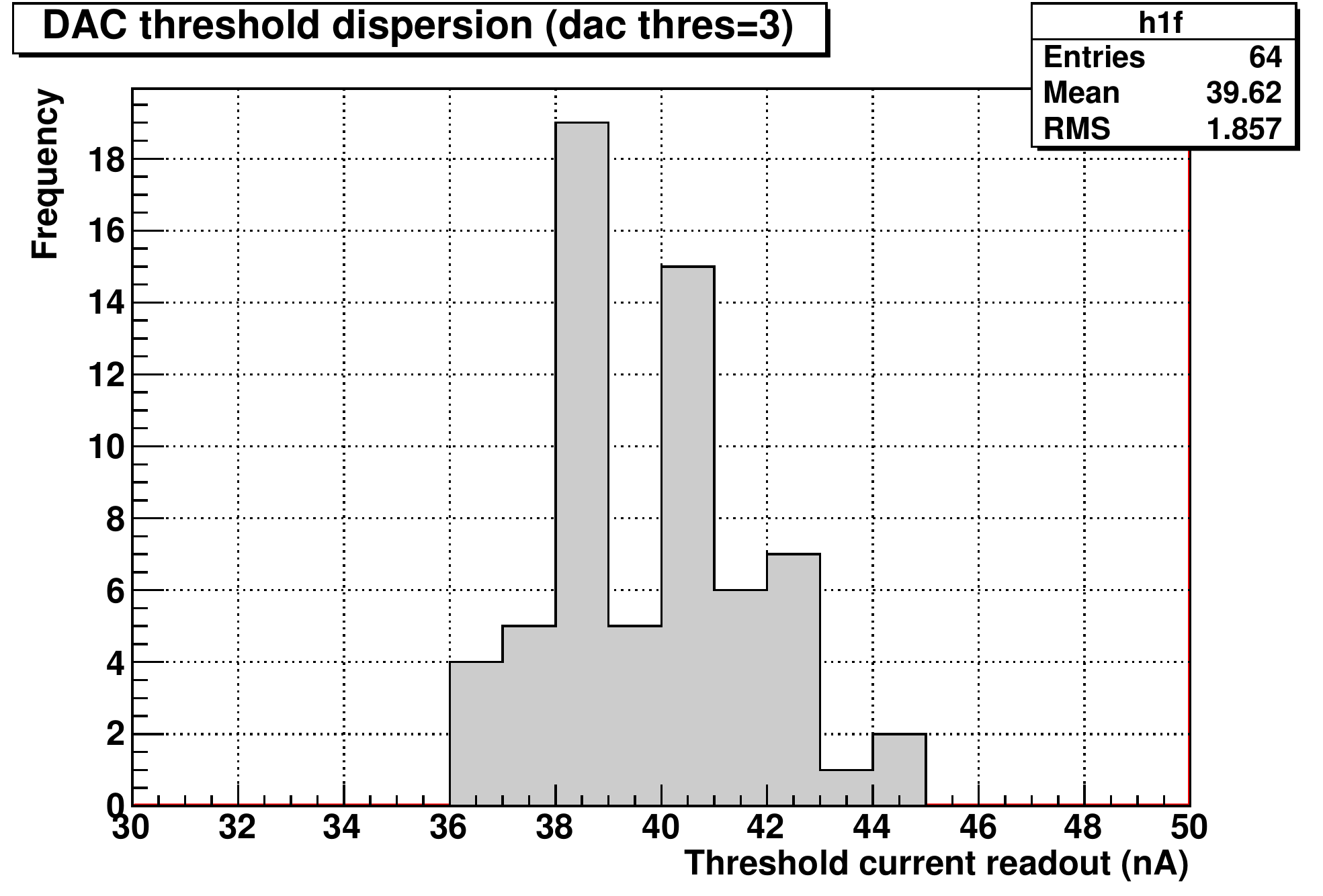}
 \includegraphics[width=0.48\textwidth]{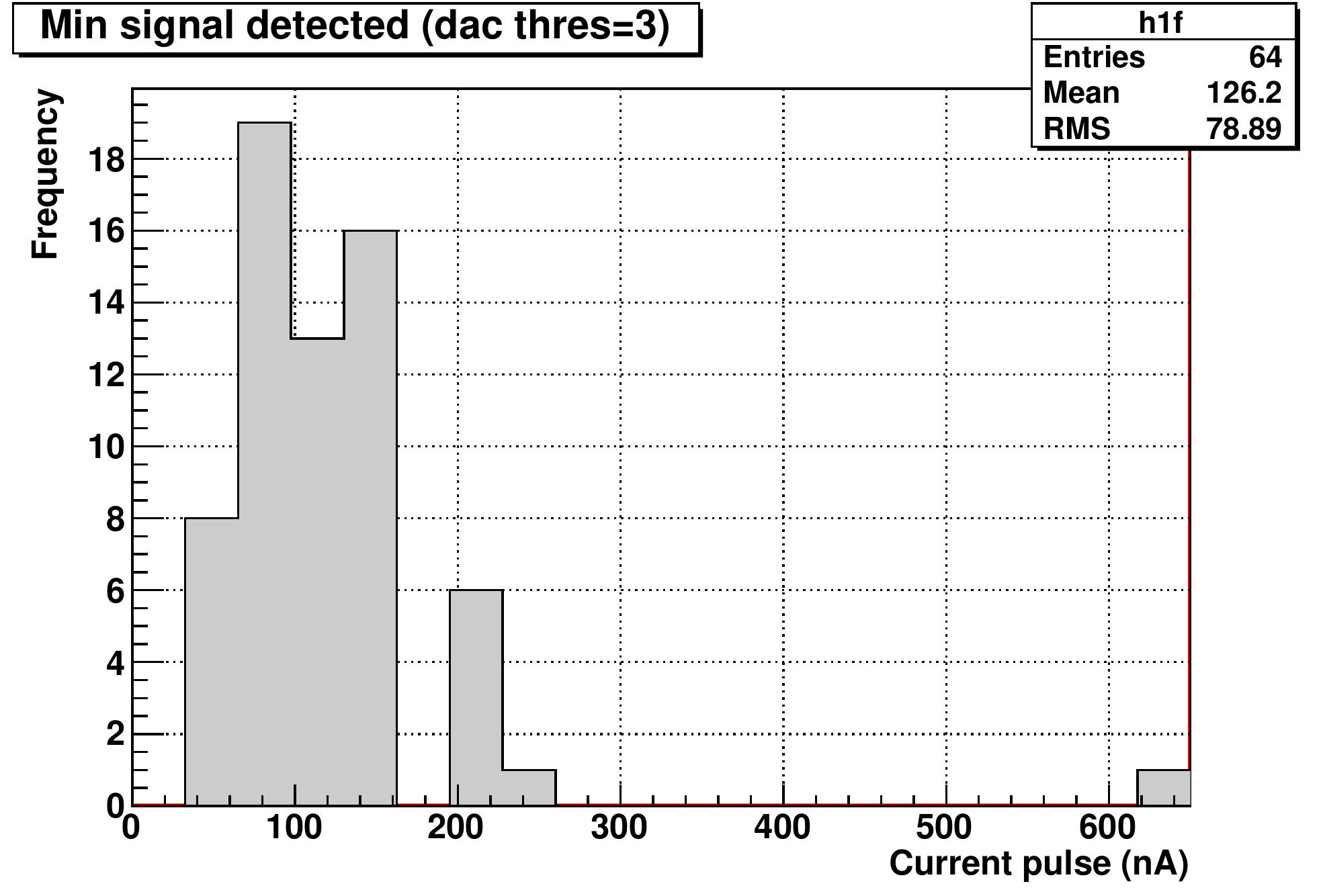}
\caption{Typical DAC threshold  and minimal threshold dispersion over one representative ASIC.}
\label{TestDiscriASIC}
\end{center}
\end{figure}

\begin{figure}[th!]
\begin{center}
  \includegraphics[width=0.5\textwidth]{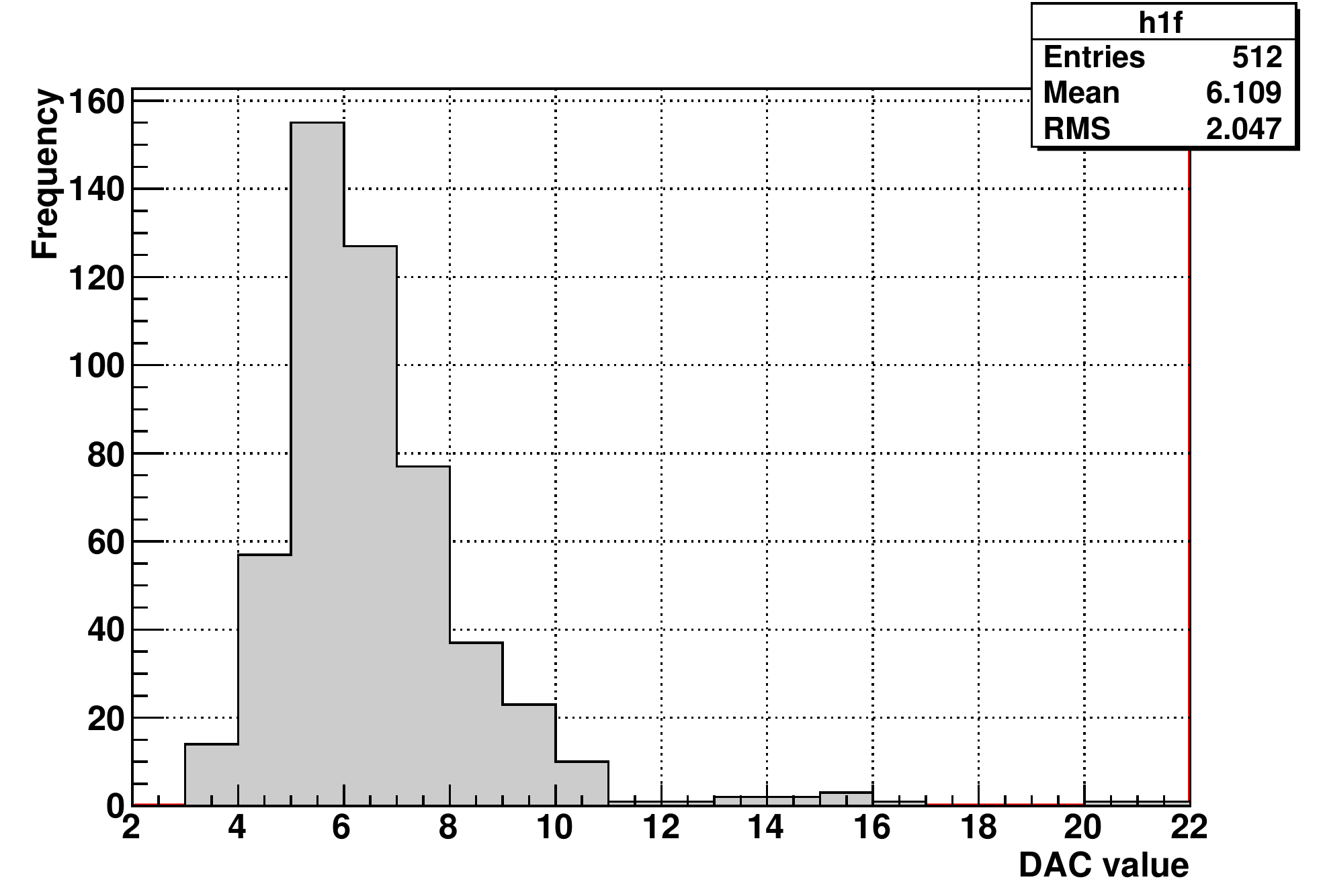}
\caption{Minimum DAC threshold dispersion above noise, in the detector.}
\label{SeuilsDACDet}
\end{center}
\end{figure}

Figure~\ref{TestDiscriASIC} shows the typical DAC threshold and minimal threshold dispersion over one representative ASIC, where the latter is determined by recording the current pulse amplitude required to trigger a channel at a fixed DAC setting. The plot shows that the DAC levels are totally homogeneous but that the effective thresholds have a few spikes remaining. This is explained by the offset correction efficiency dispersion between channels, which in turn has an influence on the minimum signal detected at a given threshold.
Figure~\ref{SeuilsDACDet} shows the minimal DAC threshold, above noise, in the detector. With the exception of a few spikes ($\sim$10), the threshold values are fairly homogeneous and allow proper operation. All the values are below maximum threshold of 413\,nA ($\rm 31 \times 200/15\ nA$).

The ASIC was fabricated in austriamicrosystems BiCMOS-SiGe 350\,nm process. It uses an effective area of $\rm 3.9\ mm \times 5.8\ mm = 22\ mm^2$ and requires a total power of 445\,mW (110\,mA at 3.3\,V, 55m\,A at 1.5\,V). 110 ASICs have been packaged in TQFP144 and tested, the production yield was about 80\%.

\section{Conclusion}
\label{SummarySec}
\begin{figure}[th!]
\begin{center}
\includegraphics[width=0.8\textwidth]{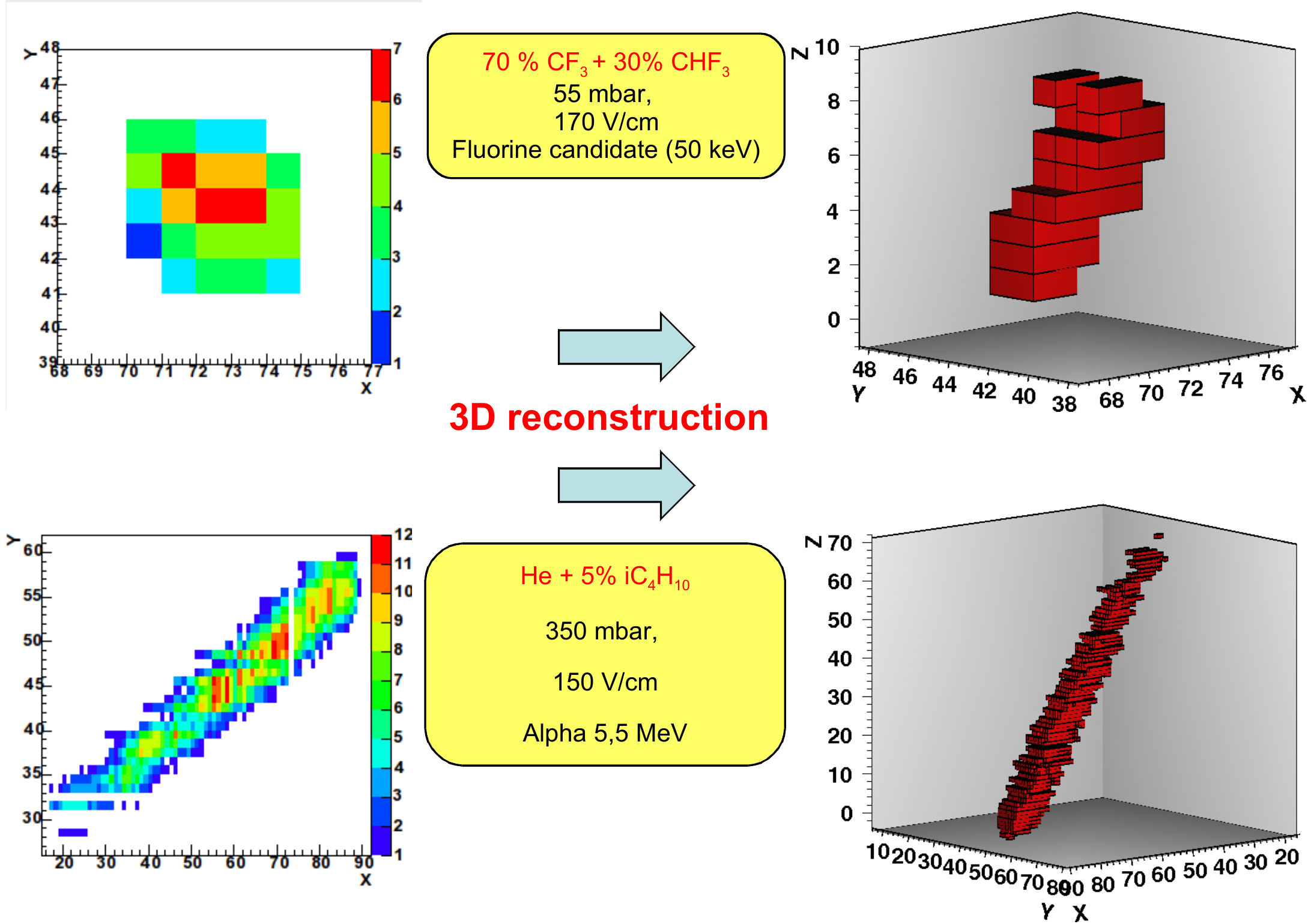}
\caption{Sample of detected recoil tracks.}
\label{traces3D}
\end{center}
\end{figure}

Eight ASICs (previous version 2) have been mounted on a specific board connected to a prototype \textmu TPC composed of $2 \times 256$ strips of pixels \cite{Bourrion64}. Figure~\ref{traces3D} shows two examples of detected recoil tracks. Position information is provided in a list of X/Y coordinates fired per time slice by the ASIC readout electronics (X-Y projection shown on the left). Offline software is used to construct the 3D display of the recoil track (3D track on the right).


\begin{thebibliography}{99}

\bibitem{Spergel} D.N. Spergel, Phys. Rev. D37 (1988) 1353  
\bibitem{Ahlen} S. Ahlen et al., Int. J. Mod. Phys. A25 (2010) 1 
\bibitem{billard.2010a} J. Billard et al.,  Phys. Lett. B691 (2010) 156-162
\bibitem{billard.2011}  J. Billard, F. Mayet and D. Santos, Phys. Rev. D83 (2011) 075002
\bibitem{billard.2010b} J. Billard, F. Mayet and D. Santos, Phys. Rev. D82 (2010) 055011


\bibitem{MIMAC} D. Santos et al.,  J.\ Phys.\ Conf.\ Ser.\   65   (2007) 012012
\bibitem{Iguaz}  F. J. Iguaz et al., 2011 JINST 6 P07002
\bibitem{Richer} J.P. Richer et al.,  Nucl.\ Instr.\ Meth.\  A620  (2010) 470-476 

\bibitem{linan}G. Linan-Cembrano, R. Del Rio-Fernandez, R. Dominguez-Castro, A. Rodriguez-Vasquez, Electronics Letter volume {\bf 33}, issue 25 (1997) p2082-p2084
\bibitem{dzahini}D. Dzahini, J. Pouxe, O. Rossetto, IEEE TNS {\bf 47} (2000) 3 839-843

\bibitem{boni}A. Boni, A. Pierazzi,  D. Vecchi, IEEE J. of Solid State Circuits. {\bf 36} (2001) 4 706-711
\bibitem{Bourrion64} O.  Bourrion et al., proceedings of TWEPP-11, Vienna, Austria, 26-30 September 2011, \href{http://arxiv.org/abs/1110.4348}{arXiv:1110.4348}

\end{thebibliography}
\end{document}